\documentclass[11pt, a04paper]{article}

\usepackage{amsmath, amssymb}
\usepackage{amsthm}

\newtheorem{theorem}{Theorem}
\newcommand{\N}{\mathbb{N}}
\newcommand{\AP}{\mathcal{AP}}
\newcommand{\SAP}{\mathcal{SAP}}
\newcommand{\EAP}{\mathcal{EAP}}

\title{Strongly Almost Periodic Sequences under\\
Finite Automata Mappings}
\author{Yuri Pritykin}
\date{\today}

\begin{document}

\maketitle

\begin{abstract}
The notion of almost periodicity nontrivially generalizes the notion
of periodicity. Strongly almost periodic sequences (=uniformly
recurrent infinite words) first appeared in the field of symbolic
dynamics, but then turned out to be interesting in connection with
computer science. The paper studies the class of eventually strongly
almost periodic sequences (i.~e., becoming strongly almost periodic
after deleting some prefix). We prove that the property of eventual
strong almost periodicity is preserved under the mappings done by
finite automata and finite transducers. The class of almost periodic
sequences includes the class of eventually strongly almost periodic
sequences. We prove this inclusion to be strict.
\end{abstract}

\section{Introduction}

Strongly almost periodic sequences (=uniformly recurrent infinite
words) were studied in the works of Morse and Hedlund
\cite{Symb01,Symb02} and of many others (for example see
\cite{Cass}). This notion first appeared in the field of symbolic
dynamics, but then turned out to be interesting in connection with
computer science.

Evidently, the class of finite automata mappings of strongly almost
periodic sequences (for definitions see below) contains the class of
eventually strongly almost periodic sequences, i.~e., becoming
strongly almost periodic after deleting some prefix. Indeed, we use
finite automaton with delay to get the sequence $a\omega$ from the
strongly almost periodic sequence $\omega$: this automaton keeps the
string $a$ in memory, first outputs this string and then outputs the
input sequence with delay $|a|$ (always remembering last $|a|$
symbols of the sequence). The main result of the article (Theorem
\ref{finiteAutomataStrong}) states the equality of the classes. In
other words, Theorem \ref{finiteAutomataStrong} says that finite
automata preserve the property of eventual strong almost
periodicity. In the last section we consider the generalized version
of finite automaton
--- finite transducer --- and prove the same statement for it.

The notion of almost periodic sequence was studied in \cite{AlPer}.
The authors prove that the class of almost periodic sequences is
also closed under finite automata mappings. It can easily be checked
that the class of almost periodic sequences contains the class of
eventually strongly almost periodic sequences. We prove this
inclusion to be strict (Theorem \ref{alPer-NotEssAlPer}).

\medskip
Let $A$ be a finite alphabet with at least two symbols. Consider the
sequences over this alphabet~--- mappings $\omega\colon\N\to A$
(where $\N=\{0, 1, 2,\dots\}$). Denote by $A^*$ the set of all
finite strings over $A$ including the empty string~$\Lambda$. If
$i\le j$ are natural, denote by $[i,j]$ the segment of natural
numbers with ends in $i$ and $j$, i.~e., the set $\{i, i+1,
i+2,\dots,j\}$. Also denote by $\omega[i,j]$ the segment of the
sequence $\omega$~--- the string
$\omega(i)\omega(i+1)\dots\omega(j)$. A segment $[i,j]$ is an
occurrence of a string $x\in A^*$ in a sequence $\omega$ if
$\omega[i,j]=x$. Denote by $|x|$ the length of the string $x$. We
imagine the sequences going horizontally from the left to the right,
so we use terms ``to the right'' or ``to the left'' to talk about
greater and smaller indices respectively.

\section{Almost periodicity}

A sequence $\omega$ is called \emph{almost periodic} if for any
string~$x$ occurring in the sequence infinitely many times there
exists a number~$l$ such that any segment of $\omega$ of length~$l$
contains at least one occurrence of~$x$. We denote the class of
these sequences by $\AP$.

A sequence $\omega$ is called \emph{strongly almost periodic} if for
any string~$x$ occurring in the sequence at least once there exists
a number~$l$ such that any segment of $\omega$ of length~$l$
contains at least one occurrence of~$x$ (and therefore $x$ occurs in
$\omega$ infinitely many times). Denote by $\SAP$ the class of these
sequences.

For convenience we introduce one additional definition: a sequence
$\omega$ is \emph{eventually strongly almost periodic} if
$\omega=a\nu$ for some $\nu\in\SAP$ and $a\in A^*$. The class of
these sequences we denote by $\EAP$.

Every eventually strongly almost periodic sequence is obviously
almost periodic. Let us show that $\EAP$ is a proper subclass of
$\AP$.

\begin{theorem}
\label{alPer-NotEssAlPer} There exists a binary sequence $\omega$
such that $\omega\in\AP$, but $\omega\notin\EAP$.
\end{theorem}
\begin{proof}
Construct a sequence of binary strings $a_0=1$, $a_1=10011$,\\
$a_2=1001101100011001001110011$, and so on, by this rule:
  $$
 a_{n+1}=a_n\bar a_n\bar a_na_na_n,
  $$
where $\bar x$ is a string obtained from $x$ by changing every 0
to 1 and vice versa. Put
  $$
 c_n=\underbrace{a_na_n\dots a_n}_{10}
  $$
and
  $$
 \omega=c_0c_1c_2c_3\dots
  $$
Prove that $\omega$ is a required one.

The length of $a_n$ is $5^n$, so the length of $c_0c_1\dots
c_{n-1}$ is $10(1+5+\dots+5^{n-1})=\frac52(5^n-1)$. By definition,
put
  $$
 l_n=\frac52(5^n-1)=|c_0c_1\dots c_{n-1}|.
  $$

Show that $\omega$ is almost periodic. Suppose $x\ne\Lambda$ occurs
in $\omega$ infinitely many times. Take $n$ such that $|x|<5^n$.
Suppose $[i,j]$ is an occurrence of $x$ in $\omega$ such that $i\ge
l_n$. By construction, for any $k$ we can consider the part of
$\omega$ starting from the position $l_k$ as a concatenation of
strings $a_k$ and $\bar a_k$. Thus by definition of $i$ the string
$x$ is a substring of one of four strings $a_na_n$, $a_n\bar a_n$,
$\bar a_na_n$, $\bar a_n\bar a_n$. Note that the string 10011
contains all strings of length two 00, 01, 10, 11, so the string
$a_{n+1}$ contains each of $a_na_n$, $a_n\bar a_n$, $\bar a_na_n$,
$\bar a_n\bar a_n$. So, $x$ is a substring of $a_{n+1}$. Similarly,
$x$ is a substring of $\bar a_{n+1}$. If the segment of length
$2|a_{n+1}|$ occurs in the sequence to the right of $l_{n+1}$, then
$a_{n+1}$ or $\bar a_{n+1}$ occurs in this segment. Hence for
$l=\frac52(5^{n+1}-1)+2\cdot5^{n+1}$ it is true, that on every
segment of length $l$ there exists an occurrence of~$x$.

Prove now that for any $n>0$ the string $c_n$ does not occur in
$\omega$ to the right of $l_{n+1}$. In this case, $c_n$ occurs
finitely many times in the sequence obtained from $\omega$ by
deleting some prefix of the length at most $l_n$, i.~e., this
sequence is not strongly almost periodic. Therefore $\omega$ is not
eventually strongly almost periodic.

Let $\nu$ be the sequence obtained from $\omega$ by deleting the
prefix of the length $l_{n+1}$. As above, for each $k$, $1\le k\le
n+1$, $\nu$ is a concatenation of strings $a_k$ and $\bar a_k$.
Assume $c_n$ occurs in $\nu$ and let $[i,j]$ be one of this
occurrences. For $n>0$ the string $c_n$ begins with $a_1$, hence
$[i,i+4]$ is an occurrence of $a_1$ in~$\nu$. We see that
$a_1=10011$ occurs in $a_1a_1=1001110011$, $a_1\bar a_1=1001101100$,
$\bar a_1a_1=0110010011$ or $\bar a_1\bar a_1=0110001100$ only in
zeroth or fifth position. Thus $5|i$, i.~e., $\nu$ and $c_n$ can be
considered as constructed of ``letters'' $a_1$ and $\bar a_1$, and
we assume that $c_n$ occurs in $\nu$. Now it is easy to prove by
induction on $m$ that $5^m|i$ for $1\leqslant m\leqslant n$, i.~e.,
we can consider $\nu$ and $c_n$ as constructed of ``letters'' $a_m$
and $\bar a_m$, and assume that $c_n$ occurs in $\nu$ (the base for
$m=1$ is already proved, but we can repeat the same argument
changing 1 and 0 to $a_m$ and $\bar a_m$ and taking into account
that $c_n$ begins with $a_m$ for each $1\le m\le n$).

Therefore we have shown that $5^n|i$, i.~e., if we consider $\nu$
and $c_n$ to be constructed of ``letters'' $a_n$ and $\bar a_n$,
then $c_n=\underbrace{a_na_n\dots a_n}_{10}$ occurs in $\nu$.
However there exists an occurrence of ``five-letter'' string
$a_{n+1}$ or $\bar a_{n+1}$ on each segment of 10 consequent
``letters'' $a_n$ and $\bar a_n$ in $\nu$, and $\bar a_n$ occurs in
this string. This is a contradiction.
\end{proof}

Moreover, it is quite easy to modify the proof and to construct
continuum of sequences in $\AP\setminus\EAP$. For example, for each
sequence $\tau\colon\N\to\{9,10\}$ we can construct $\omega_\tau$ in
the same way as in the proof of Theorem \ref{alPer-NotEssAlPer}, but
instead of $c_n$ we take
  $$
 c_n^{(\tau)}=\underbrace{a_na_n\dots a_n}_{\tau(n)}.
  $$
Obviously, all $\omega_\tau$ are different for different~$\tau$ and
hence there exists continuum of various~$\tau$.

In conclusion, we can remark that, as it is shown in \cite{AlPer},
there exists continuum of different sequences in $\SAP$ too.

\section{Finite automata mappings}

\emph{Finite automaton} is a tuple $F = \langle A, B, Q, \tilde q,
f\rangle$, where $A$ and $B$ are finite sets called input and output
alphabets respectively, $Q$ is a finite set of states, $\tilde q\in
Q$ is the initial state, and $f\colon Q\times A\to Q\times B$ is the
translation function. We say that the sequence $\langle p_n,
\beta(n)\rangle_{n=0}^\infty$, where $p_n\in Q$, $\beta(n)\in B$, is
the automaton mapping of the sequence $\alpha$ over alphabet $A$, if
$p_0=\tilde q$ and for each $n$ we have $\langle p_{n+1},
\beta(n)\rangle = f(p_n, \alpha(n))$. Thus we say that $F$ outputs
the sequence $\beta$ and denote this output by~$F(\alpha)$. If
$[i,j]$ is an occurrence of the string $x$ in the sequence~$\alpha$,
and $p_i=q$, then we say that automaton $F$ comes to this occurrence
of $x$ in the state~$q$.

In~\cite{AlPer} the following statement was proved: \emph{if $F$ is
a finite automaton, and $\omega\in\AP$, then $F(\omega)\in\AP$.}

This theorem can be expanded.

\begin{theorem} \label{finiteAutomataStrong}
If $F$ is a finite automaton, and $\omega\in\EAP$, then
$F(\omega)\in\EAP$.
\end{theorem}
\begin{proof}
Obviously, it is enough to prove the theorem for $\omega\in\SAP$, as
every eventually strongly almost periodic sequence keeps this
property after addition a prefix.

Thus let $\omega\in\SAP$. By the previous statement,
$F(\omega)\in\AP$. Suppose $F(\omega)$ is not eventually strongly
almost periodic. It means that for any natural $N$ there exists a
string that occurs in $F(\omega)$ after position $N$ and does not
occur to the right of it. Indeed, if we remove the prefix $[0,N]$
from $F(\omega)$, we do not get strongly almost periodic sequence,
hence there exists a string occurring in this sequence only finitely
many times. Then take the rightmost occurrence.

Let $[i_0,j_0]$ be the rightmost occurrence of a string $y_0$ in
$F(\omega)$. For some $l_0$ the string $x_0=\omega[i_0,j_0]$ occurs
in every segment of the length $l_0$ in~$\omega$ (by the property of
strong almost periodicity). If $F$ comes to $i_0$ in the
state~$q_0$, then $F$ never comes to righter occurrences of $x$ in
the state $q_0$ because in this case automaton outputs $y_0$
completely.

Now let $[r,s]$ be the rightmost occurrence of some string $a$ in
$F(\omega)$, where $r>i_0+l_0$. On the segment $\omega[r-l_0,r]$
there exists an occurrence $[r',s']$ of the string $x_0$. By
definition of $r$ we have $r'>i_0$. Thus assume
  $$
 i_1=r',\ j_1=s,\ x_1=\omega[i_1,j_1],\ y_1=F(\omega)[i_1,j_1].
  $$
Since $a$ does not occur in $F(\omega)$ to the right of $r$, then
$y_1$ does not occur in $F(\omega)$ to the right of $i_1$, for it
contains $a$ as a substring. Therefore if the automaton comes to the
position $i_1$ in the state $q_1$, then it never comes to righter
occurrences of $x_1$ in the state $q_1$. Since $x_1$ begins with
$\omega[r',s']=x_0$, and $r'>i_0$, we get $q_1\ne q_0$. We have
found the string $x_1$ such that automaton $F$ never comes to
occurrences of $x_1$ to the right of $i_1$ in the state $q_0$ or
$q_1$.

Let $m=|Q|$. Arguing as above, for $k<m$ we construct the strings
$x_k=\omega[i_k,j_k]$ and corresponding different states $q_k$, such
that $F$ never comes to occurrences of $x_k$ in $\omega$ to the
right of $i_k$ in the states $q_0,q_1,\dots,q_k$. For $k=m$ we have
a contradiction.
\end{proof}

\section{Finite transducers}

Let $A$ and $B$ be finite alphabets. The mapping $h\colon A^*\to
B^*$ is called \emph{homomorphism}, if for any $u,v\in A^*$ we have
$h(uv)=h(u)h(v)$. Clearly, any homomorphism is fully determined by
its values on one-character strings. Let $\omega$ be the sequence
over the alphabet $A$. By definition, put
  $$
 h(\omega)=h(\omega(0))h(\omega(1))h(\omega(2))\dots
  $$

Suppose $h\colon A^*\to B^*$ is a homomorphism, $\omega$ is an
almost periodic sequence over $A$. In \cite{AlPer} it was shown that
if $h(\omega)$ is infinite, then it is almost periodic. Thus it is
obvious, that if $\omega$ is strongly almost periodic, and
$h(\omega)$ is infinite, then $h(\omega)$ is also strongly almost
periodic. Indeed, it is enough to show that any $v$ occurring in
$h(\omega)$ occurs infinitely many times. However there is some
string $u$ occurring in $\omega$ such that $h(u)$ contains $v$, but
by the definition of strong almost periodicity $u$ occurs in
$\omega$ infinitely many times. Evidently, for $\omega\in\EAP$ we
have $h(\omega)\in\EAP$, if $h(\omega)$ is infinite.

Now we modify the definition of finite automaton, allowing it to
output any string (including the empty one) over output alphabet
reading only one character from input. This modification is called
\emph{finite transducer} (see~\cite{FinTrans}). Formally, we only
change the definition of translation function. Now it has the form
  $$
 f\colon Q\times A\to Q\times B^*.
  $$
If the sequence $\langle p_n, v_n\rangle_{n=0}^\infty$, where
$p_n\in Q$, $v_n\in B^*$, is the mapping of $\alpha$, then the
output is the sequence $v_0v_1v_2\dots$

Actually, we can decompose the mapping done by finite transducer
into two: the first one is a finite automaton mapping and another is
a homomorphism. Each of these mappings preserves the class $\AP$, so
we get the corollary: finite transducers map almost periodic
sequences to almost periodic. Similarly, by Theorem
\ref{finiteAutomataStrong} and arguments above we also get the
following statement: finite transducers map eventually strongly
almost periodic sequences to eventually strongly almost periodic (if
the output is infinite).

\bigskip {\bfseries Acknowledgements.}
The author is grateful to Alexei Semenov and Andrej Muchnik for the
help in the work and also to the participants of Kolmogorov seminar
for useful discussions.

\end{document}